\documentclass[12pt]{article}
\usepackage{amsmath,amsfonts,amssymb}
\usepackage{bbm}
\usepackage[mathcal]{euscript}
\usepackage{mathrsfs}

\newcommand{\oldcal}[1]{\CMcal{#1}}

\newcommand{\caligrafico}{\mathscr}

\newcommand{\calH}{{\caligrafico H}}

\newcommand{\calS}{{\caligrafico S}}

\newcommand{\eps}{\epsilon}

\begin{document}

\begin{center}
{\Large \textbf{Aspects of Two-Level Systems under External Time
    Dependent Fields}}

\vspace{0.2cm}

V. G. Bagrov$^\ddagger$, 
J. C. A. Barata$^\dagger$\footnote{Partially supported by CNPq.  E-mail:
  jbarata@fma.if.usp.br}, 
D. M. Gitman$^\dagger$\footnote{Partially supported by
  CNPq and FAPESP. E-mail: gitman@fma.if.usp.br}, and
W. F. Wreszinski$^\dagger$\footnote{Partially supported by CNPq. E-mail:
  wreszins@fma.if.usp.br}

$^\dagger$Universidade de S\~ao Paulo. Instituto de F\'{\i}sica\\
Caixa Postal 66318 - S\~ao Paulo - 05315 970 - SP  - Brasil

$^\ddagger$Tomsk State University and Tomsk Institute of High Current
Electronics, Russia.

\end{center}

\vspace{0.2cm}

\noindent\textbf{Abstract.} 
The dynamics of two-level systems in time-dependent backgrounds is
under consideration. We present some new exact solutions in special
backgrounds decaying in time. On the other hand, following ideas of
Feynman, Vernon and Hellwarth, we discuss in detail the possibility
to reduce the quantum dynamics to a classical Hamiltonian system.
This, in particular, opens the possibility to directly apply powerful
methods of classical mechanics (e.g. KAM methods) to study the
quantum system. Following such an approach, we draw conclusions of
relevance for ``quantum chaos'' when the external background is
periodic or quasi-periodic in time.


\section{Introduction}

\label{Sec0} \setcounter{equation}{0} \setcounter{theorem}{0}

Models of quantum two-level systems in time-dependent backgrounds are
widely used in different physical problems, with applications ranging
from condensed matter physics to quantum optics, particularly in the
semi-classical theory of the laser \cite{7}. They may, for instance,
represent the behaviour of a (frozen in space) spin $1/2$ in a
time-dependent magnetic field. In such a case, the corresponding
Schr\"odinger equation can be treated as the reduction of the Pauli
equation to the $0+1$-dimensional case.  It takes the form
(for simplicity we adopt $\hbar=1$)
\begin{equation}
i\partial_t\Psi \;= \;H(t)\Psi,
\label{EquacaodeSchroedinger}
\end{equation}
where 
$ \displaystyle \Psi = \Psi(t) =
\begin{pmatrix}
  \psi_1 (t) \\
  \psi_2 (t)
\end{pmatrix}
$,
with the quantum Hamiltonian $H(t)$  given by
\begin{equation}
H(t) \; = \; -\frac{1}{2}\vec{B}(t) \cdot \vec{\sigma},
\; = \;
- \frac{1}{2}
\begin{pmatrix}
  B_z(t) & B_x(t)-iB_y(t) \\
  {} &  \\
  {} B_x(t)+iB_y(t) & -B_z(t)
\end{pmatrix}
,
\label{1.18}
\end{equation}
$\vec{\sigma} = (\sigma_x , \; \sigma_y , \; \sigma_z)$ being
the Pauli matrices and $\vec{B}(t)= (B_x(t), \;B_y(t), \; B_z(t))$.

Equation (\ref{EquacaodeSchroedinger}) and its solutions have been
widely studied.  Our contribution in this paper is threefold: we
present a formulation of (\ref{EquacaodeSchroedinger}) in terms of
{\it classical} Hamiltonian systems in Section \ref{Sec00}, and in Section
\ref{Sec3} we present several new exact solutions for
(\ref{EquacaodeSchroedinger}) in time-dependent backgrounds which are
switched off at the time infinity. These new exact solution can be
useful to solve scattering-like problems. Finally, in Section
\ref{Sec2} we further develop the classical Hamiltonian formulation of
Section \ref{Sec00} to discuss how qualitative methods of analysis of
classical Hamiltonian systems, like the KAM method, can be used to to
shed some light on properties related to ``quantum chaos'' of
two-level systems under periodic or quasi-periodic time-dependent
interactions \cite{9}. Section \ref{Sec2} has left
several open problems which, together with applications of Section
\ref{Sec3}, will be left to further publications.
In Sections \ref{Sec3} and \ref{Sec2} we will consider the special case
\begin{equation}
  B_x(t) \; = \; -2\epsilon,  \qquad
  B_y(t) \; = \; 0 , \qquad
  B_z(t) \; = \; -2f(t),  \label{1.32}
\end{equation}
where $\epsilon$ is a constant, and $f$ (possibly after addition of a
suitable constant) decays in time.  The Schr\"odinger equation
(\ref{EquacaodeSchroedinger}) then reads
$
i\dot{\psi_{1,2}} = \pm f(t)\psi_{1,2} + \eps \psi_{2,1} 
$.
One of the basic facts we use in Section \ref{Sec3} is that the
Schr\"odinger equation above may be shown to be equivalent to the
pair of independent second order equations
\begin{equation}
  \ddot{\psi_{1,2}} + (\pm i\dot{f} + f^2 + \epsilon^2)\psi_{1,2} \; = \; 0 .
  \label{1.33}
\end{equation}
The particular Schr\"odinger equation for (\ref{1.32}) describes
two-level systems with unperturbed energy levels $\pm \eps$ ($f\equiv
0$) submitted to an external time dependent interaction $f(t)$
inducing a transition between the unperturbed eigenstates.
Alternatively, it describes a spin $1/2$ submitted to a constant
magnetic field $-2\eps$ in direction ``$x$'' and a time-dependent
magnetic field $2f(t)$ in direction ``$z$'' produced, for instance, by
a linearly (in direction ``$z$'') polarised plane wave field propagating
in direction ``$x$''. This system has been analysed by many authors in
various approximations.  For historical references, see
\cite{12,13,BarataAHP}.


\section{Classical Hamiltonian Formulation for Two-Level Systems}

\label{Sec00} \setcounter{equation}{0} \setcounter{theorem}{0}

It is known that a classical description for spinning systems is
usually related to the limit $S\to\infty$, $\hbar\to 0$ (with $\hbar
S$ constant), where $S$ is the spin value. Thus, there is a common
belief that a spin $1/2$ system is a purely quantum object. The
possibility of a pseudo-classical description of such a system does
not contradict that fact \cite{G1,G2,G3,G4,G5}.  However, as first
remarked by Feynman, Vernon and Hellwarth \cite{6}, there is a
correspondence between equation (\ref{EquacaodeSchroedinger}) and a
classical Hamiltonian system, and solutions of this mechanical system
can be used to obtain solutions of (\ref{EquacaodeSchroedinger}).
Moreover, this allows to directly apply non-perturbative methods of
classical Hamiltonian systems, like the KAM methods, to the analysis
of our time-dependent two-level systems. In Section \ref{Sec2} we will
discuss the significance of this fact to properties of two-level
system in extremal (i.e., in weak or strong coupling regime)
conditions, drawing conclusions of relevance for ``quantum chaos''
when the field is periodic or quasi-periodic.

As mentioned, the possibility to formulate
(\ref{EquacaodeSchroedinger}) in terms of a classical Hamiltonian
system has its roots in the work of Feynman, Vernon and Hellwarth
\cite{6}, who introduced an approach which is instrumental in the
semi-classical theory of the laser \cite{7}.  Consider the
Schr\"odinger equation (\ref{EquacaodeSchroedinger}) and let
\begin{equation}
  \rho(t) \; := \; |\Psi(t){\rangle } {\langle }\Psi(t)| \; = \;
\begin{pmatrix}
\psi_1 (t) \\ 
\psi_2 (t)
\end{pmatrix}
\begin{pmatrix}
  \psi_1^* (t) & \psi_2^* (t)
\end{pmatrix}
\; = \;
\begin{pmatrix}
  |\psi_1|^2 & \psi_1 \psi_2^* \\
  &  \\
  \psi_2 \psi_1^* & |\psi_2|^2
\end{pmatrix}
\label{1.24}
\end{equation}
denote the density matrix. Then $\rho$ satisfies the equation 
$ i\dot{\rho}  =  [H(t), \; \rho]$.  
Writing 
\begin{equation}
  \rho(t) \; = \; \frac{1}{2} (Q_0 {\mathbbm 1} + \vec{Q}\cdot
  \vec{\sigma}) \; = \; \frac{1}{2}
\begin{pmatrix}
  Q_0 + Q_3 & Q_1 -iQ_2 \\
  &  \\
  Q_1 +iQ_2 & Q_0 - Q_3
\end{pmatrix}
,  
\label{1.26}
\end{equation}
we have, by comparison of (\ref{1.24}) and (\ref{1.26}): 
\begin{eqnarray*}
  Q_0 & = & |\psi_1|^2 + |\psi_2|^2 \; = \; \mbox{Tr}\, \rho \; 
       = \; 1, \qquad
  Q_1  \;\ = \;\  \psi_1 \psi_2^* + \psi_2 \psi_1^*, \\
  Q_2 & = & i( \psi_1 \psi_2^* - \psi_2 \psi_1^*) , \qquad\qquad\qquad\;\;\,
  Q_3 \;\, = \;\, |\psi_1|^2 - |\psi_2|^2 .
\end{eqnarray*}
 
The equation of motion $ i\dot{\rho}  =  [H(t), \; \rho]$ yields \cite{7,6}
\begin{equation}
\dot{\vec{Q}} \; := \frac{\rm d}{{\rm d} t} \vec{Q} \; = \; 
- \vec{\Omega} \wedge \vec{Q},  
\qquad
\vec{\Omega} \; \equiv \; (B_x(t), \; B_y(t), \; B_z(t)),
\label{1.29}
\end{equation}
and the condition $\rho^2 = \rho$, which expresses that $\rho$ is a pure
state, yields 
\begin{equation}
  Q_1^2 + Q_2^2 + Q_3^2 \; = \; \vec{Q}^2 \; = \; Q_0^2 \; = \; 1 .
\label{1.31}
\end{equation}
Above henceforth the dot denotes a derivative with respect to time and
$\wedge$ denotes the vector product. 

Equations (\ref{1.29})-(\ref{1.31}) determine the wave function
completely in that $\psi_1$ and $\psi_2$ are two complex numbers, and
the phase of $\Psi$ is irrelevant. So three numbers - i.e., the vector
$\vec{Q}$ - suffice. They are the basis of a simple geometric picture
of quantum spin $1/2$ (or two-level) systems: the unit vector
$\vec{Q}(t)$ precesses around the vector $\vec{\Omega}(t)$ just like a
{\it classical} gyromagnet precesses in a magnetic field \cite{7,6}.

This observations suggests that (\ref{1.29})-(\ref{1.31}) are
associated to a classical Hamiltonian system. Let us further develop
this idea.  Let us consider the unit sphere ${\cal{S}}^2$ with the
usual angular coordinates $0 \leq \theta \leq \pi$, $0 \leq \varphi <
2\pi$, and let
\begin{equation}
\vec{\calS} \;=\; ( \calS_x, \; \calS_y, \; \calS_z)
\;=\; (\sin\theta \cos\varphi, \; \sin\theta \sin\varphi, \;
\cos\theta),
\label{dfsi}
\end{equation}
define the coordinates of a unit vector on ${\cal{S}}^2$.
Introducing 
\begin{equation}
p \; = \; \cos\theta , \qquad q = \; \varphi .
\end{equation}
as canonically conjugate variables, we may write
\begin{equation}
  \calS_x \; = \; \sqrt{1-p^2}\cos q, \quad \calS_y \; = \;
  \sqrt{1-p^2}\sin q \quad\mbox{ and } \quad \calS_z \; = \; p,
\label{SxSySz}
\end{equation}
with the usual Poisson brackets
\begin{equation}
  \left\{ \calS_x , \; \calS_y\right\} \; = \; 
    \frac{\partial \calS_x}{\partial
    q}\frac{\partial \calS_y}{\partial p} - \frac{\partial \calS_x}{\partial
    p}\frac{\partial \calS_y}{\partial q} \; = \; p \; = \; \calS_z ,
\label{PoissonBrackets}
\end{equation}
plus cyclic permutations. From (\ref{dfsi}), of course,
$
   \left( \calS_x\right)^2 + \left( \calS_y\right)^2 + \left(
    \calS_z\right)^2 =  1
$.

Let us now define in ${\cal{S}}^2$ the classical Hamiltonian
\begin{equation}
\calH^{(1)}(t) \; := \; -\vec{B}(t)\cdot \vec{\calS} . 
\label{Aga1}
\end{equation}
This describes the interaction of a classical gyromagnet with an
extremal time-dependent magnetic field $\vec{B}(t)$.
By (\ref{Aga1}) and (\ref{SxSySz}) we may write  
\begin{equation}
  \calH^{(1)} (t) \; = \;- \left[B_x (t)  \cos q + B_y (t)\sin q \right]
  \sqrt{1-p^2}  - B_z(t) p.
\label{1.20a}
\end{equation}

From (\ref{Aga1}) and (\ref{PoissonBrackets}) one sees immediately
\begin{equation}
  \dot{\vec{\calS}} \; = \; \left\{\vec{\calS}, \; \calH^{(1)}
  \right\} \; = \; - \vec{B}(t) \wedge \vec{\calS}.
\label{precS}
\end{equation}
Equation (\ref{precS})
leads to the following picture: under the time evolution defined by
$\calH^{(1)}$ the unit vector $\vec{\calS}(t)$ simply precesses
around the magnetic field vector $\vec{B}(t)$.

The important remark is that equations (\ref{1.29}) with the
parametrisation
\begin{equation}
  \vec{Q} \; = \; (\sin \theta \cos\varphi, \; \sin \theta
  \sin\varphi, \; \cos\theta)
\label{BBB}
\end{equation}
can be written in {\it classical} Hamiltonian form 
\begin{equation}
  \dot{q} \; = \; \left\{q, \; \calH^{(1)}\right\} \; = \; 
  \frac{\partial \calH^{(1)}}{\partial p}, \qquad
  \dot{p} \ =  \; \left\{p, \; \calH^{(1)}\right\} \; = \; 
  -\frac{\partial \calH^{(1)}}{\partial q},
\label{eqHamil}
\end{equation}
with $q=\varphi$, $0\leq \varphi < 2\pi$, and $p=\cos\theta$, $0 \leq
\theta \leq \pi$, and $\calH^{(1)}$ 
the classical Hamiltonian (\ref{1.20a}). 
This is immediate by comparing (\ref{precS}) with (\ref{1.29}) and
the parametrisations (\ref{dfsi}) and (\ref{BBB}).

In Section \ref{Sec2} we shall also deal with another equivalent
Hamiltonian, by the classical canonical transformation $q_1 = -p =
-\cos \theta$, $p_1 = q = \varphi$, with $0 \leq \theta \leq \pi$ and
$0 \leq \varphi \leq 2\pi$.  We again write $q_1 =q$ and $p_1=p$, so
as to keep the notation simple and put
\begin{equation}
  \calH^{(2)} (t) \; = \;- \left[B_x (t) \cos p + B_y (t)
   \sin p\right]\sqrt{1-q^2} + B_z (t) q.
\label{1.20b}
\end{equation}
The spin variables (\ref{SxSySz}) become
\begin{equation}
  \calS_x^{(2)} \; =\; \sqrt{1-q^2}\cos p, \quad \calS_y^{(2)}\; =\;
  \sqrt{1-q^2}\sin p \quad\mbox{ and } \quad \calS_z^{(2)} \; =\; -q.
\label{S2xS2yS2z}
\end{equation}
Since $\calH^{(2)}=- \vec{B}(t) \cdot \vec{\calS}^{(2)}$,
equation (\ref{precS}) reads now
\begin{equation}
  \dot{\vec{\calS}^{(2)}} \; = \; \left\{\vec{\calS}^{(2)}, \;
    \calH^{(2)} \right\} \; = \; - \vec{B}(t) \wedge
  \vec{\calS}^{(2)},
\label{precS2}
\end{equation}
again with $ \left( \calS_x^{(2)}\right)^2 + \left(
  \calS_y^{(2)}\right)^2 + \left( \calS_z^{(2)}\right)^2 = 1 $. With
the parametrisations $ p = \varphi $ and $q = -\cos\theta$, $0\leq
\varphi < 2\pi$ and $0 \leq \theta \leq \pi$, equation
(\ref{S2xS2yS2z}) becomes (as (\ref{dfsi})) the usual angular
representation of the unit vector $\vec{\calS}^{(2)}$ on the unit
sphere:
$
  \vec{\calS}^{(2)} =  (\sin \theta \cos\varphi, \; \sin \theta
  \sin\varphi, \; \cos\theta)
$.

In spite of being conceptually enlightening as discussed above, the
connection between the quantum equations (\ref{1.29}) with the
classical Hamiltonian system of (\ref{eqHamil}) does not seem to have
been applied to some of the most exciting recent developments
associated with the Hamiltonian (\ref{1.18}) for spin-$1/2$ systems in
external periodic and quasi-periodic fields \cite{9}, both in weak
coupling \cite{10} and strong coupling \cite{11,12}. This will be done
in Section \ref{Sec2}. There we show that the geometric approach
provides very interesting insights into several aspects of ``quantum
chaos'' associated to two-level systems \cite{9}.

What can we say if the external field is not periodic or
quasi-periodic? In this case some {\it exact} solutions may be found,
and in Section \ref{Sec3} we show how the geometric picture helps to
find them, having as a basis the solution for constant field. More
precisely, we consider the special case (\ref{1.32})
where $\epsilon$ is a constant, and $f$ (possibly after addition of a
suitable constant) decays in time.

\subsection{Remarks on the Semi-classical Limit of Spin Systems}

The theory of one spin (of spin quantum number $S$) or, alternatively,
a $N=2S+1$-level system, interacting with an external time-dependent
magnetic (or electric) field has always been the object of intensive
study in quantum optics and in the statistical mechanics of quantum
spin systems.

In the classical limit, $S\to \infty$, $\hbar\to
0$ with $\hbar S =1$ the spin operators $\vec{S}=(S_x, \; S_y , \;
S_z)$ satisfying the $su(2)$ commutation relations $[S_x , \; S_y ] =
i\hbar S_z$, plus cyclic permutations, converge
\cite{FawcettBracken,Lieb1} to the classical canonically conjugate
variables of a gyromagnet. More precisely
\begin{eqnarray}
  \frac{S_x}{S} \; \to \; \calS_x \; := \; \sin\theta \cos\varphi , 
\qquad
  \frac{S_y}{S} \; \to \; \calS_y \; := \; \sin\theta \sin\varphi, 
\qquad
  \frac{S_z}{S} \; \to\;  \calS_z \; := \; \cos\theta,               
\label{1.15}
\end{eqnarray}
with $0 < \theta < \pi$, $0 \leq \varphi < 2\pi$, the usual angles on the
unit sphere.

Consider now this spin in an extremal time-dependent magnetic field
$\vec{B}(t)$. The corresponding Hamiltonian 
\begin{equation}
H(\vec{S}, \; t) \; = \; - \vec{B}(t)\cdot \vec{S}
\end{equation}
satisfies, by (\ref{1.15})
\begin{equation}
\frac{H(\vec{S}, \; t)}{S} \; \to \; -\vec{B}(t)\cdot \vec{\calS} \; 
\equiv \; \calH^{(1)}(t),
\end{equation}
showing that the classical Hamiltonian
$\calH^{(1)}(t)$ is relevant {\it both} for $S\to\infty$ {\it and}
$S=1/2$! 

Classical considerations play an important role in condensed matter
physics, in particular in the theory of magnetism. There they are even
applied to the extreme quantum limit, viz., spin $1/2$, often with
remarkably good results. In order to give just one striking example,
the quantum mechanical ground state energy per spin, in the
thermodynamic limit, computed by linear spin-wave theory around the
classical ground state, is off the exact value by only $3\%$ \cite{5}.
The above features may be justified by the fact that (\ref{1.15}) is
also applicable to spin $1/2$, as we saw.  This may be surprising,
because in the spin $1/2$ case the error committed by replacing
$\vec{S}/S$ by the r.h.s. of (\ref{1.15}) is very large, but it may
explain some of the striking successes of classical considerations for
spin $S=1/2$ systems mentioned above.


\section{Exact Solutions}

\label{Sec3} \setcounter{equation}{0} \setcounter{theorem}{0}

In Section \ref{Sec2} we will learn how classical KAM methods can be
used to shed some light on the properties of some quantum systems, as
spin $1/2$ or two-level systems under the action of an external
periodic or quasi-periodic field $f$. The situation where $f$ is
non-periodic or non-quasi-periodic may be, in general, more subtle. A
surprising fact, however, is that in some situations exact solutions
can be found. Besides being interesting for their own, they may be of
relevance for the study of physical properties of the quantum systems
described, like the computation of asymptotic transition probabilities
and its large-time corrections.
 
In the present section we are going to present some exact solutions of
the equation (\ref{1.29}) or equivalently to it Eqs. (\ref{1.33}).  In
this connection, one ought to remark that the first component
$\psi_{1}(t)$ in equations (\ref{1.33}) is a solution of the
stationary one-dimensional Schr\"{o}dinger equation $
\ddot{\psi}_{1}+V\psi_{1}=0 $ with a complex potential $V$ related to
the function $f$ by a differential equation of the first order: $
V=(i\dot{f}+f^{2}+\epsilon^{2}) $. In this case, by the Schr\"odinger
equation $ i\dot{\psi_{1,2}} = \pm f(t)\psi_{1,2} + \eps \psi_{2,1} $,
the second component $\psi_{2}(t)$ can be restored from $\psi_1$ 
through
\begin{equation}
\psi_{2}\; =\; \epsilon^{-1}\left(i\partial_t - f(t) \right)\psi_{1} .
\label{restauracao} 
\end{equation}

Solutions of the one-dimensional Schr\"{o}dinger equation are
discussed in \cite{BagGi90}, whose results and considerations can be
used to find concrete functions $f$ that admit exact solutions of the
equations (\ref{1.29}) and the respective explicit solutions. Below we
present two physically interesting exact solutions of the equations
under consideration. Convergent perturbative solutions for periodic
$f$ can be found in \cite{13,BarataAHP}. 

\subsection{An Auxiliary Solution}

One can find a solution of the equations (\ref{1.29}) for $f=\mathrm{\ 
  const.}$ The vector $\mathbf{\Omega}$, given in (\ref{1.29}) is, by
(\ref{1.32})
\begin{eqnarray}
  \mathbf{\Omega } &=& -2\left( \epsilon ,\; 0,\; f\right) \;=\;-2\omega
  \left( \sin
    2\gamma ,\; 0,\; \cos 2\gamma \right) ,\;  \notag \\
  \omega &=&\sqrt{\epsilon ^{2}+f^{2}}\,,\;\;\epsilon
  \;=\;\omega \sin 2\gamma ,\;\;f\;=\;\omega \cos 2\gamma \,.
\label{3.1}
\end{eqnarray}
In the general case $ 0 \leq \gamma \leq 2\pi $, but if we restrict
ourselves to positive $\epsilon >0,$ then $ 0 \leq \gamma \leq \pi /2
$.  The general solution of the equations under consideration has the
form
\begin{align}
  \psi _{1}(t)\;& =\;+p\sin \gamma \exp \left( i\omega t\right) +q\cos
  \gamma \exp
  \left( -i\omega t\right) ,  \notag \\
  \;\psi _{2}(t)\;& =\;-p\cos \gamma \exp \left( i\omega t\right)
  +q\sin \gamma \exp \left( -i\omega t\right) \,.
\label{3.4}
\end{align}
Here $p,\;q$\ are two complex constants. Let us introduce two angles
$\varphi_{0}$ and $\psi $ by the relations 
\begin{equation}
pq^{\ast }\;=\;|pq|\exp \left( 2i\varphi _{0}\right) ,\qquad 
\psi\;=\; \omega t+\varphi_{0}\;.  
\label{3.5}
\end{equation}
Then we find 
\begin{align}
  Q_{0} \;& =\; R^{2}\;=\;\left| p\right| ^{2}+\left| q\right|
  ^{2},\qquad Q_{1} \; =\; \left( \left| q\right| ^{2}-\left| p\right|
    ^{2}\right) \sin 2\gamma -2|pq|\cos
  2\gamma \cos 2\psi ,  \notag \\
  Q_{2} \;& =\; 2|pq|\sin 2\psi , \qquad\qquad\;\; Q_{3} \; =\; \left(
    \left| q\right| ^{2}-\left| p\right|^{2}\right)
  \cos 2\gamma +2|pq|\sin 2\gamma \cos 2\psi ,  \notag \\
  |\psi _{1}|^{2} \;& =\; \left| p\right| ^{2}\sin ^{2}\gamma +
  \left|q\right|^{2}\cos ^{2}\gamma +
  |pq|\sin 2\gamma \cos 2\psi ,  \notag \\
  |\psi _{2}|^{2} \;& =\; \left| p\right| ^{2}\cos ^{2}\gamma +\left|
    q\right|^{2}\sin ^{2}\gamma -|pq|\sin 2\gamma \cos 2\psi .
\label{3.6}
\end{align}

\subsection{The First Exact Solution}

The function $f$ of the form
\begin{equation}
  f\;=\; f_{0}\tanh\tau+f_{1}\,,\qquad \tau\;=\;\frac{t}{T}
\label{3.11}
\end{equation}
admits an exact solution as will be demonstrated below. Here $f_{0}$
and $f_{1}$ are two arbitrary real constants. It is obvious that $
\lim_{t\rightarrow \pm \infty }f(t)=f_{\pm }=f_{1}\pm f_{0} $.  Thus,
at large $\left| t\right| ,$ the solution has to coincide with the
ones obtained above for constant $f_{\pm }$ . Let us introduce a new
variable $z$,
\begin{equation}
  z\;=\;\frac{1}{2}\left( 1+\tanh \tau \right) ,\qquad 0<z<1\,,
\label{3.13}
\end{equation}
and dimensionless constants 
\begin{equation}
  a\;=\;Tf_{0},\quad b\;=\;Tf_{1},\quad E\;=\;\epsilon T,\quad
  \omega _{\pm }\;=\;\sqrt{E^{2}+\left( a\pm b\right) ^{2}}\,.
\label{3.1.4}
\end{equation}
The points $z=1,0$ correspond to $t=\pm \infty $ respectively, and 
\begin{equation*}
  \frac{d}{dt} \; = \;\frac{2}{T}z\left( 1-z\right) \frac{d}{dz}, 
\qquad\quad
  \frac{d^{2}}{dt^{2}}\;  = \;\frac{4}{T^{2}}\left[ z^{2}\left(
      1-z\right)^{2} \frac{d^{2}}{dz^{2}}+z\left( 1-z\right) \left(
      1-2z\right) \frac{d}{dz} \right] \,.
\end{equation*}

We search a solution of the first equation in (\ref{1.33}) in the form 
\begin{equation}
  \psi_{1}(t)\;=\; z^{\mu }\left( 1-z\right) ^{\nu }F\left( z\right)
  \,.
\label{3.16}
\end{equation}
Taking into account that $ f=\frac{1}{T}\left( 2az+b-a\right)$,
and $ \dot{f}=\frac{4a}{T^{2}}z\left( 1-z\right) $ we obtain the
following equation for the function $F\left( z\right)$:
\begin{equation}
  z^{2}\left( 1-z\right) ^{2}\frac{d^{2}}{dz^{2}}F+z\left( 1-z\right)
  \left[ 1+2\mu -2\left( \mu +\nu +1\right) z\right]
  \frac{d}{dz}F+\Phi \left( z\right) F\;=\;0\,,
\label{3.17}
\end{equation}
where 
\begin{equation*}
  \Phi \left( z\right) \; = \;\mu ^{2}+\frac{\omega _{-}^{2}}{4}+\left(
    \nu ^{2}+
    \frac{\omega _{+}^{2}}{4}-\mu ^{2}-\frac{\omega _{-}^{2}}{4}\right) z 
   -\left( \mu +\nu +1+ia\right) \left( \mu +\nu -ia\right) z\left(
    1-z\right) \,.
\end{equation*}
Selecting $ 2\mu = i\omega _{-}$ and $2\nu = i\omega _{+}$ we
arrive at the hypergeometric equation for the function $F$ (see
\cite{GraRi} eq. 9.151). Then the general solution for the function
$\psi _{1}(t) $ has the form
\begin{equation}
  \psi _{1}(t)\;=\;c_{1}\varphi \left( \mu ,\; \nu ;\; z\right)
  +c_{2}\varphi \left( -\mu ,\; \nu ;\; z\right) \,, 
\label{3.19}
\end{equation}
where $c_{1}$ and $c_{2}$ are some complex constants, and
\begin{equation}
  \varphi \left( \mu ,\; \nu ;\; z\right)\; =\; \left(
    1-z\right)^{\nu}z^{\mu } F\left( \mu +\nu +1+ia,\;\mu +\nu
    -ia;\;1+2\mu ;\;z\right) \,.
\label{3.20}
\end{equation}
Here $F\left( \alpha ,\;\beta ;\;\gamma ;\;z\right) $ is the
hypergeometric function (see \cite{GraRi} eq. 9.100).

Taking into account (\ref{3.13}), we may write 
\begin{equation}
  z\;=\;\frac{e^{2\tau}}{1+e^{2\tau}}\,,\qquad \tau\;=\;\frac{t}{T}\,.
\label{3.20a}
\end{equation}
Thus, $\lim_{t\rightarrow-\infty}z=0\,.$ Besides,
\begin{equation}
  F\left( \alpha,\;\beta;\;\gamma;\;z=0\right) \;=\; 1\,.
\label{3.21}
\end{equation}
Then one can find the asymptotic at $t\rightarrow-\infty$,
\begin{equation}
  \psi_{1}(t)\;\approx\;
  c_{1}e^{i\omega_{-}\tau}+c_{2}e^{-i\omega_{-}\tau }\,\,.
\label{3.22}
\end{equation}
This matches with (\ref{3.4}) if we set
\begin{equation}
  c_{1}\;=\;p\sin\gamma_{-}\,,\qquad c_{2}\;=\; q\cos\gamma_{-}\,\ .
\label{3.23}
\end{equation}
The angle $\gamma_{-}$ is defined from the relations $
T\epsilon=E=\omega_{-}\sin2\gamma_{-} $ and $
Tf_{-}=\omega_{-}\cos2\gamma_{-} $.

Searching for another asymptotic at $t\rightarrow \infty $ (which
corresponds to $z\rightarrow 1$), one has to take into account that
$z=1$ is the bifurcation point of 
$F\left( \alpha ,\;\beta ;\;\gamma ;  \; z\right)$. 
Thus, to use the relation (\ref{3.21}) one has to
make the transformation $F\left( z\right) \rightarrow F\left(1-z\right)$. 
That can be done by use of the relation \cite{GraRi} 9.131.2. Then we get
\begin{equation}
  \varphi \left( \mu ,\; \nu ;\;z\right)\; =\;\bar{\varphi}\left( \mu
    ,\;\nu ;\;z\right) + \bar{\varphi}\left( \mu ,\;-\nu ;\;z\right)
  \,,
\label{3.25}
\end{equation}
where 
\begin{equation*}
  \bar{\varphi}\left( \mu ,\;\nu ;\;z\right)\; =\;\frac{\Gamma \left(
      1+2\mu \right) \Gamma \left( -2\nu \right) z^{\mu }\left(
      1-z\right) ^{\nu }}{\Gamma \left( 1+\mu -\nu +ia\right) \Gamma
    \left( \mu -\nu -ia\right) }\,.
\end{equation*}
It follows from (\ref{3.20a}) that $\lim_{t\rightarrow \infty }\left(
  1-z\right) =0.$ Taking this into account we find the asymptotic (at
$t\rightarrow \infty )$ from (\ref{3.25}), 
\begin{equation*}
  \varphi \left( \mu ,\;\nu ;\;z\right) \; \approx  \; \frac{\Gamma
    \left( 1+2\mu \right) \Gamma \left( -2\nu \right) e^{i\omega
      _{+}\tau }}{\Gamma \left(
      1+\mu -\nu +ia\right) \Gamma \left( \mu -\nu -ia\right) } 
   +\frac{\Gamma \left( 1+2\mu \right) \Gamma \left( 2\nu \right)
    e^{-i\omega _{+}\tau }}{\Gamma \left( 1+\mu +\nu +ia\right) \Gamma
    \left( \mu +\nu -ia\right) }\;.
\end{equation*}
The corresponding asymptotics for $\psi _{1}\left( t\right) $ reads 
\begin{align}
  \psi _{1}\left( t\right)\; \approx &\;\left[ \frac{\Gamma \left(
        1+2\mu \right) c_{1}}{\Gamma \left( 1+\mu +\nu +ia\right)
      \Gamma \left( \mu +\nu -ia\right)
      }
    +\frac{\Gamma \left( 1-2\mu \right) c_{2}}{\Gamma \left(
        1-\mu +\nu +ia\right) \Gamma \left( -\mu +\nu -ia\right)
      }\right] e^{i\omega _{+}\tau }
  \notag \\
   + &\; \left[ \frac{\Gamma \left( 1+2\mu \right) c_{1}}{\Gamma \left(
        1+\mu -\nu
        +ia\right) \Gamma \left( \mu -\nu -ia\right) } 
   +\frac{\Gamma \left( 1-2\mu \right) c_{2}}{\Gamma \left(
        1-\mu -\nu +ia\right) \Gamma \left( -\mu -\nu -ia\right)
      }\right] e^{-i\omega _{+}\tau }\,. \notag
\end{align}
They correspond to solutions (\ref{3.4}) with the frequency $\omega
_{+}$ in the final state, if $c_{1,2}$ obey (\ref{3.23}). Thus, the
scattering problem is solved completely without calculating the
function $\psi _{2}\left( t\right) .$ However, the latter function can
be recovered from the function $\psi _{1}\left( t\right) $ using the
second equation in (\ref{1.33}) and the formulas 9.137 of
\cite{GraRi} for the hypergeometric functions.

\subsection{Second Exact Solution}

The function $f$ of the form 
\begin{equation}
  f\;=\;\frac{f_{0}}{\cosh\tau}\,,\qquad\tau\;=\;\frac{t}{T}
\label{3.28}
\end{equation}
admits another exact solution. Here $f_{0}$ is an arbitrary real
constant.  Since $f\rightarrow0$\ at $\left| t\right|
\rightarrow\infty$, the corresponding asymptotic at $\gamma=\pi/4$
has the form (\ref{3.4}).  Introducing the variable $z$,
\begin{equation}
  z\;=\;\frac{2}{1-i\sinh\tau}\,,
\label{3.29}
\end{equation}
we find 
\begin{equation*}
  \frac{d}{dt}\;=\;\frac{z}{T}\sqrt{1-z}\frac{d}{dz}\,,\qquad
  \frac{d^{2}}{dt^{2}}\;=\; \frac{z^{2}\left( 1-z\right)
    }{T^{2}}\frac{d^{2}}{dz^{2}}+\frac{z}{2T^{2}} \left( 2-3z\right)
  \frac{d}{dz}\,.
\end{equation*}

We search a solution of the first equation in (\ref{1.33}) in the form
(\ref{3.16}) at $\mu\;=\;i\epsilon T$ and $2\nu= -Tf_{0}$.  Thus, we
find
\begin{align}
  \psi_{1}(t)\; & = \;c_{1}\varphi\left( \mu,\;\nu;\;z\right)
  +c_{2}\varphi\left(-\mu,\;\nu;\;z\right) \,,  \notag \\
  \varphi\left( \mu,\;\nu;\;z\right) \;& =\;
  \left(1-z\right)^{\nu}z^{\mu} F\left(\mu,\; \frac{1}{2}+2\nu-\mu;\;
    1+2\mu;\; z\right) \,.\,
\label{3.30}
\end{align}

As one can see, $z\rightarrow 0$ at $\left| t\right| \rightarrow
\infty$.  However, one has to be careful and consider asymptotics at
$t\rightarrow \infty $ and $t\rightarrow -\infty $ separately. Indeed,
it follows from (\ref{3.29}) that
$
  1-z=(\sinh \tau -i)/(\sinh \tau +i)
$,
and 
\begin{align}
  \left. z\right| _{t\rightarrow -\infty }& \;\approx\; -4ie^{\tau}\; =\;
  \exp \left(\tau -i\frac{\pi }{2}+\ln 4\right) \,,  \notag \\
  \left. z\right|_{t\rightarrow \infty }& \;\approx\; 4ie^{-\tau}\;=\;
  \exp \left( -\tau +i\frac{\pi }{2}+\ln 4\right) \,.
\label{3.30a}
\end{align}
Let us put $ \exp\tau =\tan \frac{\varphi }{4}$, $ 0<\varphi <2\pi$
and $1-z=\exp \varphi $.  Then, $t\rightarrow -\infty \Longrightarrow
\varphi \rightarrow 0;\;t\rightarrow \infty \Longrightarrow \varphi
\rightarrow 2\pi $, and we have $ \lim_{t\rightarrow -\infty }\arg
\left( 1-z\right) =0 $ and $ \lim_{t\rightarrow \infty }\arg \left(
  1-z\right) =2\pi $.  Taking this into account and remembering
(\ref{3.21}), (\ref{3.30a}), we get at $t\rightarrow -\infty $
\begin{equation}
  \psi _{1}\left( t\right) \;\approx \;c_{1}\exp (\chi _{1})+c_{2}\exp
  \left( -\chi _{1}\right) \,,\qquad \chi _{1}\;=\; i\epsilon
  t+\frac{\pi }{2}\epsilon T+i\epsilon T\ln 4\;.
\label{6.32}
\end{equation}
The corresponding asymptotic at $\;t\rightarrow \infty $ has the form 
\begin{equation}
  \psi _{1}\left( t\right) \;\approx\; e^{-i\pi f_{0}T}\left[
    c_{1}\exp (-\chi _{2})+c_{2}\exp \left( \chi _{2}\right) \right]
  \,,\qquad \chi _{2} \;= \; i\epsilon t+\frac{\pi }{2}\epsilon
  T-i\epsilon T\ln 4\;.
\label{6.33}
\end{equation}
As one can see there is a complete correspondence with (\ref{3.4}). At
$t\rightarrow \infty $ we may observe an exchange of the coefficients
and an additional phase appears.


\section{``Quantum Chaos'' in Two-Level Systems}

\label{Sec2} \setcounter{equation}{0} \setcounter{theorem}{0}

The problem of ``quantum chaos'' has attracted a lot of attention in
recent times (see \cite{9} and references quoted therein).  We will
now focus on it from the point of view of the classical Hamiltonian
system provided by (\ref{eqHamil}) for the Hamiltonian (\ref{1.20b}),
describing the two-level systems discussed above with periodic or
quasi-periodic time-dependent interactions.

Let us consider  the situation where
\begin{equation}
  B_x(t) \; = \; 2f(t),  \qquad
  B_y(t) \; = \; 0 , \qquad
  B_z(t) \; = \; -2\epsilon,  \label{1.32n} 
\end{equation}
we get from (\ref{1.18}) the quantum Hamiltonian
\begin{equation}
  H^{(1)}(t) \; = \; \epsilon \sigma_z - f(t)\sigma_x . 
\label{2.5}
\end{equation}
This is the most usual form of the Hamiltonian of a time-dependent
two-level system: $\epsilon$ is the energy difference of the
(unperturbed) levels in a two-level atomic system, and $-f(t)\sigma_x$
is the interaction with an external electro-magnetic field in a
two-level approximation \cite{7}.
By (\ref{1.20b}), the corresponding classical Hamiltonian is
\begin{equation}
  \oldcal{H}_1 \; = \; -2f(t) \sqrt{1-q^2} \cos p -2\eps q .
\label{2.3r}
\end{equation}

Rotation of $\pi/2$ around the $y$-axis leads from (\ref{2.5}) to 
\begin{equation}
  H^{(2)}(t) \; = \; \epsilon \sigma_x + f(t)\sigma_z ,
\label{2.5V2}
\end{equation}
which corresponds to
\begin{equation}
  B_x(t) \; = \; -2\eps,  \qquad
  B_y(t) \; = \; 0 , \qquad
  B_z(t) \; = \; -2f(t),  \label{1.32r} 
\end{equation}
in (\ref{1.18}).
The classical Hamiltonian (\ref{1.20b}) becomes
\begin{equation}
  \oldcal{H}_2 \; = \; 2\epsilon \sqrt{1-q^2} \cos p - 2f(t) q .
\label{2.3}
\end{equation}
In both cases, the situation where $\epsilon$ ``small'' is called the
strong-coupling case \cite{11,12} and the situation where
$f$ is ``small'' is called the weak-coupling case. We will analyse
both separately. We will consider (\ref{2.3}) for the strong-coupling
regime and (\ref{2.3r}) for the weak-coupling regime.

We now consider $f$ periodic with frequency $\omega$: 
\begin{equation}
  f \; = \; f(\omega t).  
\label{2.7}
\end{equation}
We are led, by Howland's method in classical mechanics (see \cite{9}
or \cite{CFKS}, chapter 7.4), to consider the autonomous Hamiltonians
corresponding to (\ref{2.3r}) and (\ref{2.3}). Roughly speaking, this
method allows to transform a non-autonomous Hamiltonian $H(q, \; p, \;
\omega t)$ into an autonomous one by treating $\omega t$ as a
coordinate $\theta$ with a corresponding canonically conjugate
momentum $I$. The associated autonomous Hamiltonian is $K (q, \; p, \;
\theta, \; I) = H(q, \; p, \; \theta) + \omega I$ and one easily
checks the equivalence of the Hamilton equations for both.

Let us denote by $\oldcal{K}_1$
and $\oldcal{K}_2$ the autonomous Hamiltonians corresponding to
(\ref{2.3r}) and (\ref{2.3}), respectively.  For (\ref{2.3r}) we get
\begin{equation}
  \oldcal{K}_1 \; = \; \oldcal{H}_1^0 + \epsilon \oldcal{V}_1, 
\quad \mbox{ where } \quad
  \oldcal{H}_1^0 \; = \; -2f(\theta) \sqrt{1-q^2}\cos p + \omega I
  \quad \mbox{ and } \quad 
\oldcal{V}_1 \; = \; -2 q,
\label{2.8r}
\end{equation}
defined on the Cartesian product phase-space $\Pi_1 \times \Pi_2$, where
\begin{eqnarray*}
  \Pi_1 & = & \{ (q, \; p); \; -1 \leq q \leq 1 ; \; 0 \leq p < 2\pi
  \mbox{ with }
  2\pi \mbox{ and } 0 \mbox{ identified}\} , \\
  \Pi_2 & = & \{ (\theta, \; I); \; 0 \leq \theta < 2\pi \mbox{ with }
  2\pi \mbox{ and } 0 \mbox{ identified};\; -\infty < I < \infty\}.
\end{eqnarray*}
Above, $I$ is the variable canonically conjugate to the angle
$\theta$, with $\dot{\theta} = \frac{\partial \oldcal{K}_1}{\partial
  I} = \omega $.  On the other hand, for (\ref{2.3}) we get
\begin{equation}
  \oldcal{K}_2 \; = \; \oldcal{H}_2^0 + \epsilon \oldcal{V}_2,
\quad \mbox{ where } \quad
  \oldcal{H}_2^0 \; = \; -2f(\theta) q + \omega I 
\quad \mbox{ and }  \quad 
\oldcal{V}_2 \; = \; 2\sqrt{1-q^2} \cos p.
\label{2.8}
\end{equation}
Again, $I$ is the variable canonically conjugate to the angle
$\theta$, with $\dot{\theta} = \frac{\partial \oldcal{K}_2}{\partial
  I} = \omega $.

The important observation now is that $\oldcal{H}_2^0$ is integrable.
In fact, $\oldcal{H}_2^0$ and $q$ are two independent constants of
motion in involution.  $\oldcal{K}_2$ is, however, not integrable and
for $\epsilon$ ``small'' $\oldcal{K}_2$ is, by (\ref{2.8}), a small
perturbation about an integrable Hamiltonian. Hence, KAM methods are
applicable \cite{9} to the analysis of the Hamiltonian system
associated to $\oldcal{K}_2$ and to the corresponding quantum spin
$1/2$ or two-level system.  Before we discuss the consequences of this
fact below let us look at the situation for the weak coupling regime.

For weak coupling it is more natural to write $\eps \equiv \omega_0$
and $f \equiv \tilde{\eps}\,\tilde{f}$ for $\tilde{\eps}$ ``small''.
Equations (\ref{2.5}) and (\ref{2.5V2}) become $ \tilde{H}^{(1)} =
\omega_0\sigma_z - \tilde{\epsilon}\, \tilde{f}(t)\sigma_x $ and
$\tilde{H}^{(2)}(t) = \omega_0\sigma_x
+\tilde{\epsilon}\,\tilde{f}(t)\sigma_z $, respectively.  The
classical autonomous Hamiltonians $\oldcal{K}_1$ and $\oldcal{K}_2$
become
\begin{equation}
  \tilde{\oldcal{K}}_1 \; = \; \tilde{\oldcal{H}}_1^0 + 
    \tilde{\epsilon}\, \tilde{\oldcal{V}}_1, 
\quad \mbox{ where } \quad
  \tilde{\oldcal{H}}_1^0 \; = \; -2\omega_0 q + \omega I 
\quad \mbox{ and } \quad 
\tilde{\oldcal{V}}_1 \; = \; -2 \tilde{f}(\theta)
  \sqrt{1-q^2}\cos p
\label{2.11}
\end{equation}
and
\begin{equation}
  \tilde{\oldcal{K}}_2 \; = \; \tilde{\oldcal{H}}_2^0 +
  \tilde{\epsilon}\, \tilde{\oldcal{V}}_2,
\quad \mbox{ where } \quad
 \tilde{\oldcal{H}}_2^0 \; = \; 2\omega_0 \sqrt{1-q^2} \cos p + \omega I 
\quad \mbox{ and } \quad 
\tilde{\oldcal{V}}_2 \; = \;
  -2\tilde{f}(\theta) q.
\end{equation}

Now, $\tilde{\oldcal{H}}_1^0$ is integrable, since $q$ and $I$ or $q$
and $\tilde{\oldcal{H}}_1^0$ are independent constants of the motion
in involution. $\tilde{K}_1 $, however, is not integrable, and again,
by (\ref{2.11}), is a small perturbation about an integrable
Hamiltonian. Therefore, KAM methods are again applicable. Notice that
in (\ref{2.11}), with $q=\calS_z=I_1$ and $I=I_2$, one has
$\tilde{\oldcal{H}}_1^0 = \tilde{\oldcal{H}}_1^0(I_1, \; I_2)$ which
is the standard form of integrable $\tilde{\oldcal{H}}_1^0$.

Several remarks already follow from this description. Firstly
$\oldcal{K}_2$ and $\tilde{\oldcal{K}}_1$ are non-integrable even in
the periodic case, which lends further insight into the nontrivial
character of the (quantum) perturbation theory developed in
\cite{13,BarataAHP}.  Secondly, the complete equivalence of the
classical dynamics described by (\ref{2.8}) or (\ref{2.11}) to the
quantum evolution throws further light into properties of the quantum
system, as we now discuss briefly.

In the periodic case (\ref{2.7}), $\oldcal{K}_2$ and
$\tilde{\oldcal{K}}_1$ (given by (\ref{2.8}) and (\ref{2.11}),
respectively) are Hamiltonians of a system of two degrees of freedom.
They are thus expected to exhibit an Aubry-Mather transition \cite{9},
at a certain critical $\epsilon_c$, which may correspond to the first
avoided crossing. The ingenious method of \cite{14}, which combines
the KAM transformation with a specific treatment of resonances and
pushes the convergence radius of the classical perturbation expansion
up to $|\epsilon| = \epsilon_c$ (or $|\tilde{\epsilon}| = \epsilon_c$)
may, if applicable to the present classical model, be translated
exactly to the quantum case, with interesting implications to a
modified Rayleigh-Schr\"odinger perturbation theory for the Floquet
eigenvalues of the quantum system.

As a final interesting insight provided by the classical description,
consider the case of quasi-periodic $ f(t) = f(\omega_1 t , \;
\omega_2 t) $ with two incommensurate frequencies \cite{9,12}. 
In cases (\ref{2.8}) and (\ref{2.11}) we are led to
three-degrees of freedom Hamiltonians
\begin{equation}
  \oldcal{K}_2 \; = \; \oldcal{H}^0_2 + \epsilon \oldcal{V}_2 ,  
\quad \mbox{ where } \quad
  \oldcal{H}^0_2 \; = \; -2f(\vec{\theta})q + \vec{\omega}\cdot \vec{I}
\quad \mbox{ and } \quad
  \oldcal{V}_2 \; = \; 2\sqrt{1 - q^2 }\cos p
\end{equation}
and 
\begin{equation}
  \tilde{\oldcal{K}_1}\;  = \; \tilde{\oldcal{H}}_1^0 +
  \tilde{\epsilon}\, \tilde{\oldcal{V}}_1 , 
\quad \mbox{ where } \quad
  \tilde{\oldcal{H}}^0_1 \; = \; -2\omega_0 q + \vec{\omega}\cdot
  \vec{I}  
\quad \mbox{ and } \quad
\tilde{\oldcal{V}}_1 \; = \;
  -2\tilde{f}(\vec{\theta})\sqrt{1 - q^2 }\cos p ,
\end{equation}
respectively, with $\vec{\theta} := (\theta_1 , \; \theta_2)$,
$\vec{I} := (I_1 , \; I_2)$, $\vec{\omega} := (\omega_1 , \;
\omega_2)$. It has, in general, quite different critical properties
from the two-degree of freedom case! \cite{15}. This may be a clue to
the nature of the differences between the periodic and the
quasi-periodic case. Although the quasi-energy spectra are dense pure
point in both cases \cite{9,16}, there are basic differences in the
nature of the perturbative series (without secular terms
\cite{12,13,BarataAHP}) in the coupling constant $\epsilon$: in
contrast to the periodic case, in the quasi-periodic case the series
is not, for reasons explained in \cite{12}, expected to define an
analytic function in any circle $|\epsilon| \leq \epsilon_0$ (however
small $\epsilon_0$) for any values of the frequencies and coefficients
of the Fourier expansion of $f$ (which are supposed to be $O(1)$ with
respect to $\epsilon$).


\section{Some Final Remarks}

\label{Sec4} \setcounter{equation}{0} \setcounter{theorem}{0}

For certain Hamiltonians which are at most quadratic in coordinates
and momenta obeying the Heisenberg-Weyl algebra (flat phase space),
there exist different explicit expressions for the basic quantum
mechanical quantities in terms of classical solutions \cite{2}. As an
example, we mention the well-known expression of the transition
amplitude via the van-Vleck determinant \cite{1}. 

In the case of compact phase space considered in this paper, there are
two semi-classical approaches: the WKB theory for spin, due to van
Hemmen and S\"{u}t\H{o} \cite{HemmenSuto}, and the path integral
formalism (see \cite{Schulman}, chap. 23 and references given there),
but the connection with classical dynamics is not established for any
spin quantum number, but only in the classical limit $\hbar \to 0$, $S
\to \infty$ with $\hbar S =1$.

The phase space path integral for spin has also been employed, notably
in \cite{Schilling}, who uses the Villain approximation. In this
context, but along different lines, we have shown that the classical
Hamiltonian (\ref{1.20a}) (or (\ref{1.20b})) is relevant to both the
classical and extreme quantum (spin $1/2$) limits of the Hamiltonian
of a quantum spin in an external magnetic field, one of whose
components is a time-dependent function $f$.




\begin{thebibliography}{99}
  
  
\bibitem{7} H. M. Nussenzveig. {\it Introduction to Quantum Optics}.
  (Gordon and Breach, New York, 1973).
  
\bibitem{FawcettBracken} R. J. B. Fawcett and A. J. Bracken. J. Phys.
  A \textbf{24}, 2743 (1991).
  
\bibitem{Lieb1} E. H. Lieb. Commun. Math. Phys. 31, 327-340 (1973).


\bibitem{5} P. W. Anderson. Phys. Rev. \textbf{86}, 694 (1952). For
  the classical spin wave theory in the ferromagnetic case, see N.
  Straumann.  {\it Klassiche Mechanik}. (Springer Verlag, Berlin,
  1987).

\bibitem{6} R. P. Feynman, F. L. Vernon Jr. and R. W. Hellwarth. J.
  Appl.  Phys. \textbf{28}, 49 (1957).

\bibitem{9} H. R. Jauslin. {\it Stability and Chaos in Classical and
    Quantum Hamiltonian Systems}. In P. L. Garrido and J. Marro, eds.
  II Granada Lectures in Computational Physics. pp. 107-172. (World
  Scientific, Singapore, 1993).
  
\bibitem{10} P. M. Bleher, H. R. Jauslin and J. L. Lebowitz. J. Stat.
  Phys.  \textbf{68}, 271 (1992).

\bibitem{11}  W. F. Wreszinski and S. Casmeridis. J. Stat. Phys. 
\textbf{90}, 1061 (1998).

\bibitem{12} J. C. A. Barata. Rev. Math.  Phys. \textbf{12}, 25-64
  (2000).
  
\bibitem{CFKS} H. L. Cycon, R. G. Froese, W. Kirsch and B. Simon.
  \textit{ Schr\"{o}dinger Operators.} (Springer Verlag, Berlin,
  1987).

\bibitem{13} J. C. A. Barata and W. F. Wreszinski. Phys. Rev. Lett.
  \textbf{84}, 2112-2115 (2000).

\bibitem{BarataAHP} J. C. A. Barata. To appear in Annales Henri
  Poincar\'{e}.
  
\bibitem{14} C. Chandre, M. Govin and H. R. Jauslin. Phys. Rev. Lett.
  \textbf{79}, 3881 (1997); Phys. Rev. E \textbf{57}, 1536 (1998).

\bibitem{15} C. Chandre and H. R. Jauslin. Phys. Rev. Lett.
  \textbf{81}, 5125 (1998).

\bibitem{16} R. Krikorian. Ann. Sci. Ecole Norm. S. 32: (2) 187-240
  Mar-Apr (1999) and R. Krikorian,  {\it Reducibility of skew-product
    systems with values in compact groups}.  Asterisque 259, III-+
  (1999).
  
\bibitem{BagGi90} V. G. Bagrov and D. M. Gitman, \emph{Exact Solutions
    of Relativistic Wave Equations}, (Kluwer, Dordrecht, Boston,
  London 1990).

\bibitem{G1} F. A. Berezin, M.S. Marinov.  Ann. Phys.  {\bf 104}, 336
  (1977).

\bibitem{G2} L. Brink, S. Deser, B. Zumino, P. di Vecchia and P. Howe.
  Phys. Lett. {\bf B64}, 435 (1976).
  
\bibitem{G3} L. Brink, P. di Vecchia and P. Howe. Nucl. Phys. {\bf
    B118}, 76 (1977).
  
\bibitem{G4} R. Casalbuoni. Nuovo Cimento {\bf A33}, 115, 389 (1976).
  
\bibitem{G5} A. Barducci, R. Casalbuoni and L. Lusanna. Nuovo Cimento
  {\bf A35}, 377 (1976).

\bibitem{GraRi}  I. S. Gradshtein, I. W. Ryzhik. \textit{Table of
    Integrals, Series and Products}. (Academic Press, New York, 1994).
  
\bibitem{2} V. Dodonov and V. Man'ko, \emph{Invariants and the
    Evolution of Nonstationary Quantum Systems}, Proceedings of
  Lebedev Physics Institute, vol. 183 (M.A. Markov, ed.), (Nauka,
  Moscow, 1987) [translated by Nova Science, Commack, New York, 1989].

\bibitem{1} J. H.  Van Vleck, Proc. Nat. Acad. Sci. USA \textbf{14},
  178 (1928); C.  Morette, Phys. Rev. \textbf{81}, 848 (1951).

\bibitem{HemmenSuto} J. L. van Hemmen and A. S\"ut\H{o}. Physica {\bf
    141B}, 37 (1986). 

\bibitem{Schulman} L. S. Schulman. {\it Techniques and Applications
    of Path Integration}, (Wiley, New York, 1981).

\bibitem{Schilling} M. Enz and R. Schilling. J. Phys. {\bf C19}, 1765
  (1986).  


 
\end{thebibliography}
\end{document}